%% file: 00_Tritium22_ViMA.tex
\documentclass{style/nseJournal}
\usepackage{siunitx}
\usepackage{ngerman}
\usepackage{nicefrac}
\usepackage{lineno}
\usepackage{upgreek}
\usepackage{pifont}
\begin{document}

\title{ViMA – the spinning rotor gauge to measure the viscosity of tritium between 77 and 300 K} 

\addAuthor{\correspondingAuthor{Johanna Wydra}}{a}
\correspondingEmail{johanna.wydra@kit.edu}
\addAuthor{{Alexander} Marsteller}{a,b}
\addAuthor{Robin Gr{\"o}{\ss}le}{a}
\addAuthor{Florian Priester}{a}
\addAuthor{Michael Sturm}{a}

\addAffiliation{a}{Institute for Astroparticle Physics, Tritium Laboratory Karlsruhe(IAP-TLK),\\ Karlsruhe Institute of Technology (KIT), \\ Hermann-von-Helmholtz-Platz 1 \\ Eggenstein-Leopoldshafen \\ 76344 \\ Baden-W{\"u}rttemberg \\ Germany}
\addAffiliation{b}{Center for Experimental Nuclear Physics and Astrophysics, \\ Department of Physics, University of Washington (UW), \\ Seattle \\ 98195 \\ Washington \\ United States of America}

\addKeyword{tritium}
\addKeyword{viscosity}
\addKeyword{TLK - Tritium Laboratory Karlsruhe}
\addKeyword{spinning rotor gauge}
\addKeyword{cryogenic setup}

\titlePage
\selectlanguage{english}
\begin{abstract}
\input{01_abstract}
\end{abstract}


\input{02_introduction}

\input{03_measurement_procedure}

\input{05_design_constraints}

\input{07_design}

\input{08_example_commissioning_measurements}

\input{09_summary}

\pagebreak

\input{10_acknowledgements}

\pagebreak
\bibliographystyle{style/ans_js}                                                                           
\bibliography{bibliography}

\end{document}

%% file: 01_abstract.tex
Experimental values for the viscosity of the radioactive hydrogen isotope tritium (T$_2$) are currently unavailable in literature. 
The value of this material property over a wide temperature range is of interest for applications in the field of fusion, neutrino physics, as well as to test ab initio calculations. 
As a radioactive gas, tritium requires careful experiment design to ensure safe and environmental contamination free measurements.
In this contribution, we present a spinning rotor gauge based, tritium compatible design of a gas viscosity measurement apparatus (ViMA) capable of covering the temperature range from \SIrange{80}{300}{\kelvin}.

%% file: 02_introduction.tex
\section{Introduction}
\label{sec:Intro}
The viscosity is one of the fundamental material properties of gases, needed for gas dynamic calculations.
Therefore, the viscosity of many gases has been measured over a wide temperature range \cite{assael2018, Rietveld1957,Itterbeek1938,Itterbeek1940,KESTIN19591033,Kestin1980}.
For many simple systems like noble gases\cite{Hurly2000} or molecular hydrogen\cite{mehl2010,Schaefer2010}, ab initio calculations have been performed.
Simple molecules which have not been covered widely due to their radioactivity are the tritiated isotopologues of hydrogen.
Ab initio calculations exist \cite{Song2016}, but since they have been carried out using a classical approach and neglect quantum effects, they are only suitable for temperatures of \SI{300}{\kelvin} and above.
To the best of the authors' knowledge, no experimental values for the viscosity of tritiated hydrogen molecules exist in published literature.

The cryogenic region of viscosity covered by neither theoretical calculations nor experiment is of interest for several applications such as the closed tritium cycle design, development and operation for nuclear fusion and experimental neutrino physics.
The Karlsruhe Tritium Neutrino Experiment (KATRIN) is one such experiment which aims to measure the electron antineutrino mass, using the electron spectrum of the tritium $\upbeta$-decay.
The KATRIN collaboration has published a new experimental upper limit on the neutrino mass of \SI{0.8}{\electronvolt} (\SI{90}{\percent C.L.}) \cite{katrin2022direct}.
In this experiment, tritium is circulated through a \SI{10}{\meter} long windowless gaseous tritium source (WGTS) at a temperature of \SI{80}{\kelvin}.
Viscosity is the central parameter to model the density profile of tritium inside the WGTS and is therefore a systematic effect for the KATRIN Experiment. 
In fusion fuel cycle the viscosity of tritium is needed for example to simulate the tokamak exhaust process, isotope separation systems based on gas chromatography and others.

In order to measure the viscosity of tritium down to temperatures relevant to the KATRIN experiment, we have built a viscosity measurement apparatus (ViMA) based on a spinning rotor gauge.
In this paper, we describe an apparatus design capable of measuring gas viscosities from \SIrange{77}{300}{\kelvin} in a manner that is compatible with the material requirements for later use with tritium.

%% file: 03_measurement_procedure.tex
\section{theory of the spinning rotor gauge (SRG)}
\label{sec:Theory}
Whereas the spinning rotor gauge (SRG) normally is used as a pressure gauge for ultra high vacuum, where its measurements are independent of the viscosity of the surrounding gas, it can also be used at higher pressures, between \SIrange{20}{2000}{\pascal}, to measure the viscosity of the gas, if the pressure is measured separately. 
In both cases, the normalized deceleration rate of the rotor is measured in dependence of the pressure and the temperature. 
The underlying effect used here, is the damping of the rotor, caused by the collisions of the surrounding gas with the SRG rotor. The torque $D$ normalized to the frequency $\Omega$ of the rotor can be written as
\begin{equation}
    \frac{D}{\Omega} = I\cdot \frac{\nicefrac{\mathrm{d}\Omega}{\mathrm{d}t}}{\Omega}\,, \label{eq:Torque/frequency}
\end{equation}
with $I$ being the moment of inertia of a sphere. By inverting the normalized torque, the formula for measuring the viscosity with the SRG, which is derived in \cite{Tekasakul1996,Bentz1997,Loyalka1996}, reads as follows:
\begin{equation}
    \frac{1}{\nicefrac{D}{\Omega}} = \frac{1}{A \cdot C_0 \mu} + \frac{1}{p} \cdot B\,, \label{eq:srgvisc}
\end{equation}
where $A$ and $B$ are constants for a given temperature $T$, $p$ is the pressure, $\mu$ is the viscosity and $C_0$ is a calibration factor (compare to \cite{Wydra2022}).
Experimentally, the viscosity is derived using \autoref{eq:srgvisc} by fitting a linear function to the taken data and extracting $\mu$ from the y-axis intersection. 
As for the measurement procedure, there are different methods to acquire the necessary data for the fit, which are currently under investigation in terms of reproducibility, systematic effects and handling.

%% file: 05_design_constraints.tex
\section{design constraints}
\label{sec:constraints}
Since the setup is planned to operate with tritium, there are some constraints. 
The whole system needs to be built according to the technical terms of delivery and acceptance (TLA) and the best engineering practice \cite{Welte2015}.
The technical design pressure range is limited to \SI{90}{\kilo\pascal}, so that in case of leakage, tritium release against atmospheric pressure is minimized. 
The whole primary system, which is in contact with tritium, needs to be fully built out of metal including the sealing gaskets. 
In addition, the system needs to fit into a glove box, leaving enough space to handle it using the gloves.
Furthermore, there are specific needs, to guarantee most accurate measurements of the viscosity.
The thermal stabilization has to be better than \SI{1}{\kelvin\per\hour}, to exclude any thermal drift inside the system during one measurement. 
The temperature gradient inside the measurement cell should not exceed \SI{1}{\kelvin}.
The pressure range inside the measurement cell needs to run from \SIrange{100}{2000}{\pascal} at room temperature and from \SIrange{20}{500}{\pascal} at liquid nitrogen temperature. 
The temperature is limited on the low end to \SI{77}{\kelvin} by the use of liquid nitrogen as a coolant and on the high side to \SI{330}{\kelvin} by the maximum operating temperature of the SRG.
Lower temperatures, down to \SIrange{10}{20}{\kelvin} where the saturation vapor pressure becomes limiting, are desirable, but were not targeted in this setup due to the associated increase in system complexity of going below liquid nitrogen temperatures.

%% file: 07_design.tex
\section{Design}
\label{sec:design}

\begin{figure}[t]
    \begin{center}
    \includegraphics[width=\textwidth]{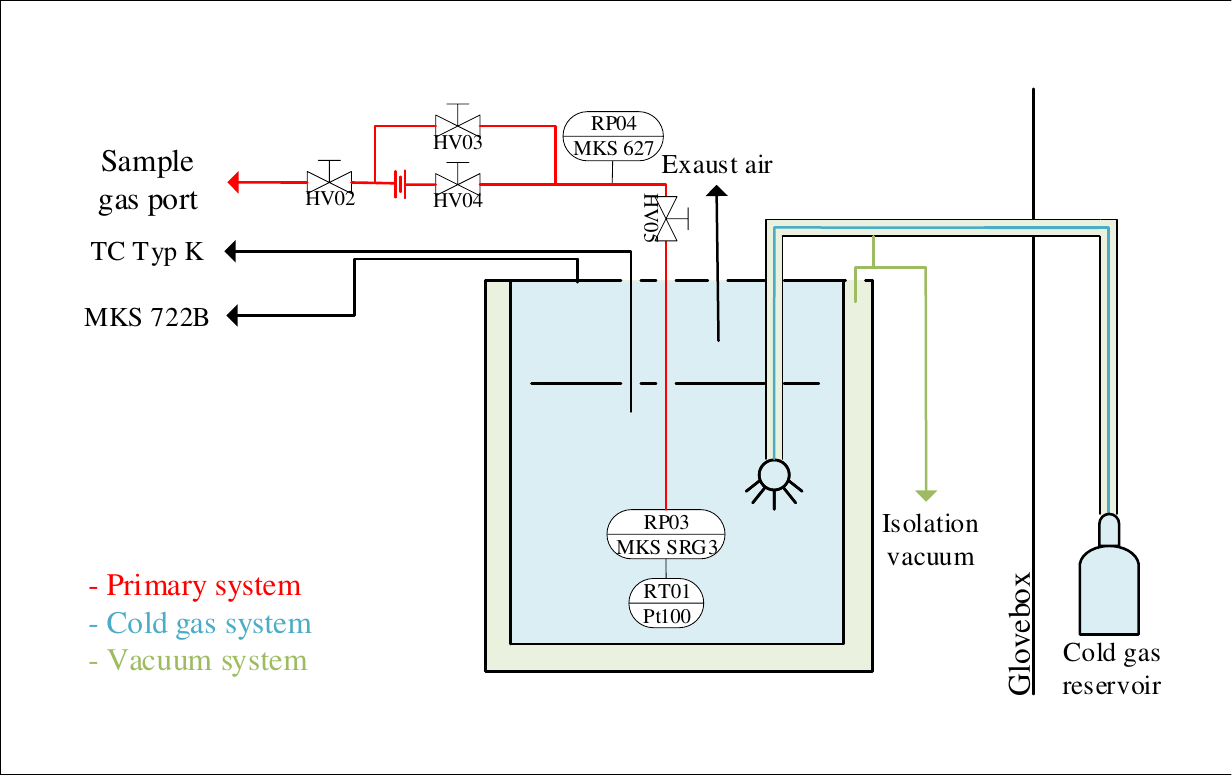}
    \end{center}
    \caption{\textbf{Flow pattern of Cryo-ViMA.} The primary system, meaning all parts in contact with tritium, are colored in red. The cold gas is marked in blue and the isolation vacuum is colored in green.}
    \label{fig:FlowPattern}
\end{figure}

\begin{figure}[t]
    \begin{center}
    \includegraphics[width=\textwidth]{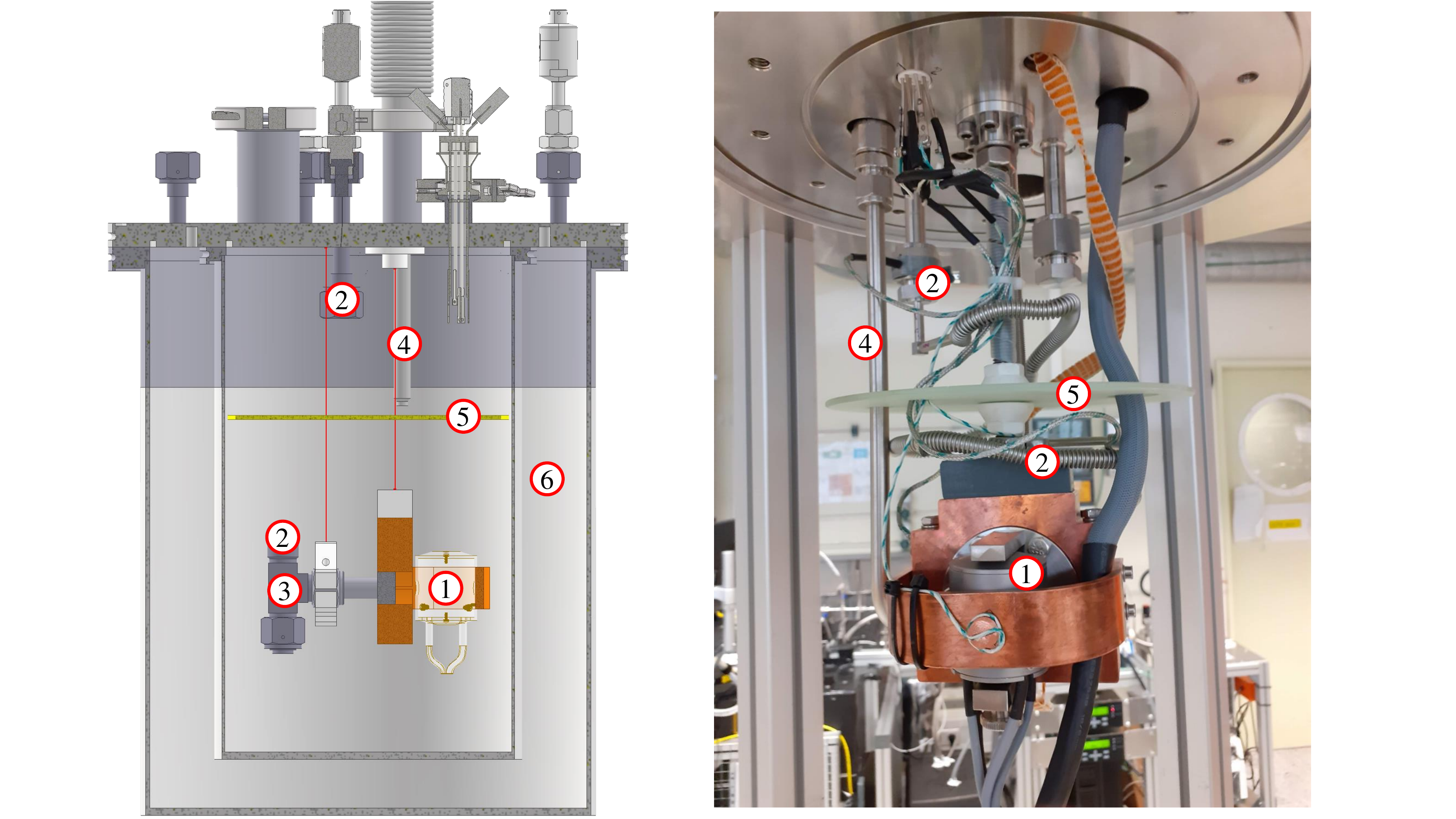}
    \end{center}
    \caption{\textbf{CAD model and photograph of Cryo-ViMA.} 
    Shown are the SRG head \ding{172} connected to that sample gas tube \ding{173} (bellow connection not shown in CAD), which also contains the sample gas temperature sensor \ding{174}.
    Cold nitrogen gas from the cold gas system is injected via a nozzle \ding{175}.
    A diffusion plate \ding{176} improves temperature stability and cooling efficency.
    Surrounding these components is a vacuum insulated Dewar \ding{177}. }
    \label{fig:Cryo_ViMA_Assembly}
\end{figure}
After proof-of-principle setups demonstrated the feasibility of the approach, the new and refined Cryogenic viscosity measurement apparatus (Cryo-ViMA) will enable us to measure the viscosity of tritium, while fulfilling all the constraints listed above. 
An annotated CAD model and photograph of the setup are shown in \autoref{fig:Cryo_ViMA_Assembly}.
To have a thermal insulation against the environment, the SRG, together with five temperature sensors (thermocouple type K), is built up inside a cryostat, consisting of two stainless steel barrels with insulation vacuum in between. 
This cryostat is cooled inside with nitrogen gas by a dedicated cold gas system. 
The cold gas system can be controlled in the whole temperature range between \SIrange{80}{300}{\kelvin} and is stabilized to \SI{2}{\kelvin\per\hour}.
At the time of writing this paper, the thermal stabilization of the cryostat is not sufficient enough for the planned measurement procedure, which is why the procedure has been changed as described in \autoref{sec:commissioning}.
Since the sample gas is stored at room temperature, a pre-cooling is needed. This is realised with a long, thin-walled bellows tube, which is installed as a loop through the whole cryostat, before it connects to the SRG thimble.
The tritium compatibility is guaranteed by only using stainless steel in the primary system, which is proved to hold pressures up to \SI{500}{\kilo\pascal}.
This qualification for high pressures is necessary to ensure that even if the system is filled with the maximum operating pressure of \SI{90}{\kilo\pascal} at \SI{77}{\kelvin}, a warm-up to room temperature does not lead to leakage or structural damage of the system.
In this fashion, even in the case of failure of the cooling system or operator mistakes, safe enclosure of tritium in the primary system can be guaranteed.

%% file: 08_example_commissioning_measurements.tex
\section{Example commissioning measurements}
\label{sec:commissioning}
\begin{figure}[tbh]
    \begin{center}
    \includegraphics[width=\textwidth]{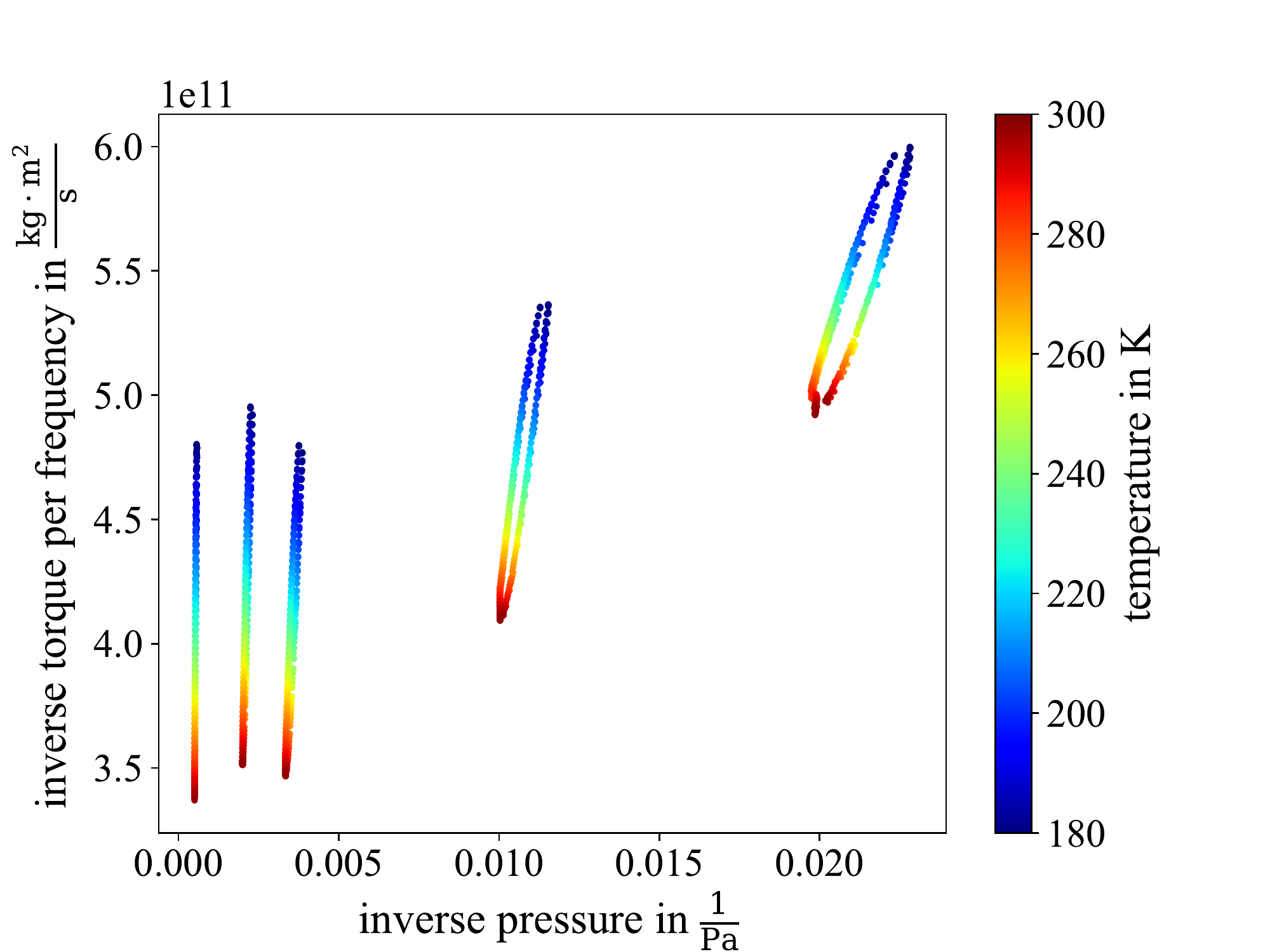}
    \end{center}
    \caption{\textbf{Example commissioning measurement.} Shown are the thermal cycle measurements at different pressure setpoints. The color indicates the temperature between \SIrange{180}{300}{\kelvin}. At low pressures (right side of the graph) a sligt hysteresis can be seen, indicating, that some adjustments still have to be made for the cycling-speed.}
    \label{fig:Measurements}
\end{figure}
The measurements described in \cite{Wydra2022} are conducted at constant temperature by changing the pressure inside the SRG. 
The temperature of Cryo-ViMA can only be stabilized to \SI{2}{\kelvin\per\hour}, so measurements with constant temperature are not feasible for high precision results.
With Cryo-ViMA, the measurements are conducted by fixing the absolute particle number inside the relevant volume, by choosing a pressure set-point between \SIrange{10}{2000}{\pascal} and closing the volume by closing valves HV03 and HV04 (compare to \autoref{fig:FlowPattern}).
This way, the pressure inside the SRG is still monitored.
Afterwards the temperature of the cold gas system is ramped from room temperature down to \SI{77}{\kelvin} and up again. 
One such thermal cycle takes about one day and is repeated for different pressure set-points. 
The result can be seen in \autoref{fig:Measurements}, where every loop corresponds to one thermal cycle. 
From here the viscosity can be calculated by picking measured values for one temperature and fitting \autoref{eq:srgvisc}. 

%% file: 09_summary.tex
\section{Summary}
\label{sec:conclusion}
The measurement of gas viscosity using rotating bodies is an established method in literature \cite{Loyalka1996,shimokawa2016}, and SRGs have been used for this at room temperature in the past \cite{Loyalka1996,Bentz1997,Tekasakul1996}.
Our efforts have been focused on extending these prior works with the goal of achieving measurements at cryogenic temperatures and featuring tritium compatibility.
In this contribution, we have presented a setup that enables the viscosity measurement of gases in the range from \SIrange{77}{300}{\kelvin}.
We have taken care in our design and material selection to achieve safe enclosure of the sample gas, allowing us to use the setup to measure the viscosity of the radioactive hydrogen isotope tritium.
In general, the setup we have presented in this contribution can be used to measure the viscosity of hazardous (radioactive, corrosive, toxic, ...) gases, making it applicable outside of our particular field of research as well.

%% file: 10_acknowledgements.tex
\section*{Acknowledgments}
We acknowledge the support of the Helmholtz Association (HGF), Germany, the German Ministry for Education and Research BMBF (05A20VK3) and the DFG graduate school KSETA, Germany (GSC 1085).